\documentclass[twocolumn,showpacs,preprintnumbers,amsmath,amssymb,aps,prl]{revtex4}

\usepackage{graphicx}% Include figure files

\begin{document}

\preprint{}

\title{Phase Instability as a Source of Modal Dynamics in Semiconductor Lasers}
% Force line breaks with \\

\author{L. Gil$^{1,2}$} \email{lionel.gil@inln.cnrs.fr} \author{G.L. Lippi$^{1,2}$}
\address{$^1$ Institut Non Lin\'eaire de Nice, Universit\'e de Nice-Sophia Antipolis, Valbonne, France}
\address{$^2$ CNRS, UMR 7735, 1361 Route des Lucioles, F-06560, France}

\begin{abstract}
Thanks to a new derivation of the fundamental equations governing multimode dynamics for a semiconductor laser near its threshold, we identify regimes of existence of a pure phase instability (and of a mixed phase-amplitude turbulence regime) which give an alternative satisfactory interpretation of the deterministic multimode dynamics observed in some devices. The existence of intrinsic noise generated by the phase instability reconciles in the same description the deterministic and random features of the semiconductor dynamics.
\end{abstract}

\date{\today}% It is always \today, today,

%semiconductor laser, General laser theory, Dynamical laser instabilities; noisy laser behavior
\maketitle
%%%%%%%%%%%%%%%%%%%%%%%%%%%%%%%%%%%%%%%%%%%%%%%%%%%%%%%%%%%%%%%%%%%%%%%%%%%%%%%%%%%%%%%%%

Random modal switching has been recognized as the dominant feature in semiconductor lasers oscillating on two or more longitudinal modes, driven either by mode partition -- where the total laser power fluctuates among several coexisting longitudinal modes \cite{Linke1985,Ohtsu1989} -- or by mode-hopping -- where only one mode at a time is emitting \cite{Ohtsu1989,Ohtsu1986}. Theoretical modeling has been based on phenomenological multimode rate equations \cite{Copeland1983,Henry1984,Wentworth1990}, where noise plays a crucial role in forcing mode-switching.

The existence of a {\it deterministic} modal switching dynamics, characterized by a periodic oscillation of the optical frequency among a few adjacent modes, has been experimentally found in Multiple-Quantum-Well (MQW) lasers~\cite{Yacomotti2004,Furfaro2004}, where the intensity of each mode vanishes periodically while keeping the sum of the intensities constant, thanks to an appropriate lag between alternating modes. Such dynamics are not limited to MQW lasers but have also been reported in multimode Quantum Dot lasers~\cite{Tanguy2006}.  Theoretical explanations have been based on modal cross-talk through Four-Wave-Mixing (FWM)~~\cite{Yacomotti2004}, Cross- (CS) and Self-Saturation (SS)~\cite{Ahmed2003}.  Noise, omnipresent but fundamentally incompatible with the deterministic dynamics, becomes -- in this description -- the motor of the dynamics if introduced as a source of fluctuations not at the modal but at the total electric field level, its projection onto the modal amplitudes preserving the phase correlations among the modes  \cite{Ahmed2002}.

A new approach, based on two Complex Ginzburg-Landau Equations (CGLE) for the slowly varying amplitudes of two counter-propagating optical fields coupled to an equation for the carrier population dynamics \cite{Serrat2006}, describes the full electric field, thus preserving the entire phase information and including all nonlinear effects otherwise modeled by nonlinear expansions (FWM, CS, SS, etc.).   This model correctly reproduces the anti-correlated oscillations of the modal intensities and the stationarity of the incoherent sum (i.e., neglecting the fast oscillations due to inter-mode beating), but, contrary to common belief, it cannot describe the destabilization of the monochromatic solution near threshold (cf. later discussion).
 
Building on previous work and aware of the systematic errors introduced by the usual approximations (Rotating Wave, Slowly Varying Envelope and direct Adiabatic Elimination of the medium's polarization) \cite{Gil2011}, we use a standard multi-scale analysis \cite{Iooss} which provides an accurate description of the dynamics of the coupled electromagnetic (e.m.) field ($E$) and carrier density ($N$) near the laser threshold.  We thereby obtain a more sophisticated and complete model, whose highly technical and lengthy derivation will be published elsewhere~\cite{inpreparation}.  Our model, which only superficially resembles~\cite{Serrat2006}, contains among other features a new control parameter ($\beta$) -- whose definition parallels that of the   $\alpha$-parameter -- responsible for the appearance of a true phase-unstable regime~\cite{BenjaminFeir1967} in a semiconductor laser (the phase instability is the equivalent, for dissipative systems, of self-focussing for conservative systems).

One should not confuse the occurrence of a pure phase instability, i.e. constant laser intensity, with the constant obtained when performing the incoherent sum of the modal intensities~\cite{Serrat2006}, since the inter-mode beats can produce an amplitude dynamics undetected by the incoherent sum.  In addition, phase dynamics may also produce coupled phase-amplitude turbulence~\cite{GLPhasePortrait} which cannot be captured by less sophisticated models.  Although at the present stage no experimental measurements are available to discriminate among the two mechanisms (as the beat frequencies were outside the experimental detection bandwidths~\cite{Yacomotti2004,Furfaro2004,Tanguy2006}), the appearance of a new destabilization mechanism opens very interesting questions on the role of noise on the dynamics. Indeed, the phase dynamics playing the role of a substantial noise source~\cite{Manneville}, it is natural to wonder about its relative weight compared to that of the usual noise sources~\cite{Henry1986,Osinski1987}.

The main features of our model \cite{inpreparation} can be summarized as follows.  The electric field $E$, polarized along $x$ and propagating along $z$, satisfies
\begin{equation}
\partial_{tt}E+{{1}\over{\epsilon_{0}}}\partial_{tt}P=c^2 \partial_{zz}E-\sigma \partial_{t}E \, ,
\label{EquationOndes}
\end{equation}
where $P(t,z)$ is the dielectric polarization, $\epsilon_0$ and $c$ are the dielectric constant and the speed of light in vacuum, respectively, and $\sigma$ represents losses. The carrier density obeys
\begin{equation}
\partial_{t} N=\gamma \left(N_{p}-N\right)+D \partial_{zz}N+{{2}\over{\hbar \omega_{c}}}E \partial_{t}P \, ,
\label{EquationPopulation}
\end{equation}
where $\gamma$ is the carrier's relaxation constant, $N_{p}$ the pump parameter, $D$ the diffusion constant, $\hbar$ Planck's constant, and $\omega_{c}$ is the oscillation frequency. The Fourier transforms of $E$ and $P$ are related through the susceptibility $\chi(\omega,N)$:
\begin{equation}
\widehat{P}(\omega)=\epsilon_{0} \chi(\omega,N) \widehat{E}(\omega) \, .
\label{DefinitionChi}
\end{equation}

Close to transparency three independent slow characteristic time scales appear and are associated with: (i) the electric field amplitude growth rate (related to the distance $\Delta N$ from transparency); (ii) the population inversion time constant ($\gamma^{-1}$); (iii) $\Gamma^{-1}$ $\propto$ ${{\partial \chi}\over{\partial \omega}} \huge{\vert}_{\omega_{c},N_{pc}}$ which characterizes the susceptibility's frequency dependence near resonance.

An adiabatic elimination based on the eigendirections~\cite{Oppo1986} of the vector field offers better predictions than those of the direct adiabatic elimination.  However, an even more accurate description is obtained with the help of a standard codimension two analysis~\cite{Coullet1999}, i.e. by inspecting the evolution of the electric field (thus also of the polarization, eq.~(\ref{DefinitionChi})) on a time scale comparable to that of the population in a neighborhood of the bifurcation. The exponent ($x$$>$$0$), measuring the ratio between the two characteristic times ${{\Delta N}\over{N_{pc}}}=\left({{\gamma}\over{\omega_{c}}}\right)^{x}
$, is well-known to lead to a slightly different final system of equations and a different dynamics depending on the scaling choice~\cite{Coullet1999}. In order to estimate the best value for $x$, we analyze the different scales intervening in the problem:  those fixed by the physical parameters give ${{\gamma}\over{\omega_{c}}}\simeq 10^{-6}$, ${{\Gamma}\over{\omega_{c}}} \simeq 10^{-3}$, ${{D k_{c}^{2}}\over{\omega_{c}}} \simeq 10^{-3}$ and ${{\sigma}\over{\omega_{c}}}\simeq 10^{-3}$.  Close to the bifurcation, the scaling law for the e.m. field amplitude can be estimated from eq.~(\ref{EquationPopulation}) by neglecting the diffusion term for the population, considering equilibrium, and extracting a relation between $N$ and $E$:
\begin{equation}
\label{e-scale}
\gamma \left(N_{p}-N\right)\propto {{2}\over{\hbar \omega_{c}}} E\partial_{t}P
\Longrightarrow  {{\gamma N_{pc} \hbar}\over{\epsilon_{0}}} {{dN}\over{N_{pc}}} \propto  {{\sigma}\over{\omega_{c}}} E^2 \, ,
\end{equation}
with $k_c$ and $\omega_c$ threshold wavevector and frequency, respectively, and using $Im\left({{\partial \chi}\over{\partial \omega}}\huge{\vert}_{\omega_{c},N_{pc}}\right)=0$, which offers a straightforward relation between $P$ and $E$.  From eq.~(\ref{e-scale}) we immediately obtain
$
E \propto \sqrt{{{\gamma \hbar N_{pc}}\over{\epsilon_{0}}}} \left({{\gamma}\over{\omega_{c}}}\right)^{{{x}\over{2}}-{{1}\over{4}}}
$
which requires $x > {{1}\over{2}}$ for the $E$ field to be small.  At the same time the smallest allowed values for $x$ imply a large distance from threshold and longer analytical computations.  Testing for several values of $x$ gives the usual answer for {\it regular} physical problems (i.e., with no additional symmetry breaks):  the general shape of the result is conserved, differences appear only in the quantitative values of the coefficients.  Thus, on the basis of this check we have proceedeed with $ x = {{3}\over{5}}$, which represents a good compromise between accuracy and computing effort.

We introduce the following scalings and definitions:
\begin{equation}
\begin{array}{lll}
\gamma = \omega_{c} \epsilon^{20}, & N_{p}  = N_{pc}\left(1+ \widetilde{\mu} \epsilon^{12} \right)  & \chi=\chi_{r}+i\chi_{i}
\cr
\sigma = \omega_{c} \epsilon^{10} \widetilde{\sigma}, & \Gamma = \omega_{c} \epsilon^{10} \widetilde{\Gamma},
&
D = \epsilon^{10} {{\omega_{c}}\over{k_{c}^{2}}} \widetilde{D}
\end{array}
\end{equation}
as well as the dimensionless partial derivatives
\begin{equation}
\begin{array}{lllll}
\chi_{\omega}=\left(\gamma \omega_{c} \right)^{{1}\over{2}} {{\partial \chi}\over{\partial \omega}}
& &
\chi_{\omega\omega}=\left(\gamma \omega_{c} \right) {{\partial^2 \chi}\over{\partial \omega^2}}
& &
\chi_{N}=N_{pc} {{\partial \chi}\over{\partial N}} \, ,
\end{array}
\end{equation}
where we stipulate that $\chi$ will appear instead of $\chi(\omega_{c},N_{pc})$, whenever $\chi$ or any of its derivatives are evaluated at $\omega_{c}, N_{pc}$.  The unusual value of the exponent for the small parameter $\epsilon$ is chosen to avoid manipulating fractional exponents~\cite{notsosmall}.

As customary, slow space ($Z$) and time ($T$) coordinates, together with order parameters ($S$ and $F$), are defined as:
\begin{equation}
\begin{array}{lccl}
N &= &N_{pc}&\left(1+\epsilon^{12} S+\epsilon^{13} N_{13}+...\right)
\cr
E &= &\sqrt{{{\gamma N_{pc} \hbar}\over{\epsilon_{0}}}}&\left(\epsilon^1 F+\epsilon^2 E_{2}+\epsilon^{3} E_{3}+...\right)
\cr
P &= &\epsilon_{0} \sqrt{{{\gamma N_{pc} \hbar}\over{\epsilon_{0}}}}&\left(\epsilon^1 P_{1}+\epsilon^2 P_{2}+\epsilon^{3} P_{3}+...\right)
\cr
\partial_{z} &=&k_{c} & \left(\partial_{z_{0}}+\underbrace{\epsilon^{10} \partial_{z_{10}}+\epsilon^{11} \partial_{z_{11}}+\epsilon^{12} \partial_{z_{12}}+...}_{\partial_{Z}=\epsilon^{10} \partial_{\tilde{Z}}}\right)
\cr
\partial_{t} &=&\omega_{c} & \left(\partial_{t_{0}}+\underbrace{\epsilon^{20} \partial_{t_{10}}+\epsilon^{21} \partial_{t_{21}}+\epsilon^{22} \partial_{t_{22}}+...}_{\partial_{T}=\epsilon^{20} \partial_{\tilde{T}}}\right)
\end{array}
\end{equation}
The previous expansions are solutions of eqs.~(\ref{EquationOndes},\ref{EquationPopulation},\ref{DefinitionChi}) provided that:
\begin{subequations}\label{EquationsFinales}
\begin{align}
\label{modela}
\partial_{\widetilde{T}}F&=-V \partial_{\widetilde{Z}} F+\epsilon^{2}c_{0}SF 
\\ \nonumber
&+\epsilon^{10}c_{1}\partial_{\widetilde{Z}\widetilde{Z}}F+\epsilon^{12}c_{2}S \partial_{\widetilde{Z}}F \, ,
\\
\label{modelb}
\partial_{\widetilde{T}}S&=\widetilde{\mu}-S-4\widetilde{\sigma}\vert F \vert^2+\epsilon^{10}\widetilde{D}\partial_{\widetilde{Z}\widetilde{Z}}S \, ,
\end{align}
\end{subequations}
where
\begin{equation}
\begin{array}{ll}
V={{2c^2k_{c}^2}\over{\omega_{c}^2 \chi_{\omega}}}
&
c_{0}={{\chi_{iN}}\over{\chi_{\omega}}} \left(1-i \alpha\right)
\cr
c_{1}={{-V^2\chi_{i\omega\omega}}\over{2\chi_{\omega}}}\left(1-i \beta \right)
&
c_{2}={{-i \chi_{\omega\omega} V c_{0} }\over{\chi_{\omega}}}+{{\chi_{\omega N} V}\over{\chi_{\omega}}}
\end{array}
\end{equation}
with $\mathcal{R}e\{c_0\}$ and $\mathcal{R}e\{c_1\}$$>0$ (by construction), $\alpha$ $=$ $\chi_{rN} / \chi_{iN}$ is the usual alpha-factor and $\beta$ a new, real function defined as
\begin{equation}
\beta={{\chi_{\omega}}\over{V \chi_{i\omega\omega}}}\left({{\chi_{r\omega\omega}V}\over{\chi_{\omega}}}-1\right) \, .
\end{equation}
The first line in eq.~(\ref{modela}) contains the usual slowly varying envelope terms for class B semiconductor lasers \cite{Spinelli,NJP}, while the second line contains the smaller new terms. The term containing the complex coefficient $c_{1}$ represents diffusion (as in \cite{Serrat2006}) but also diffraction of the electric field, while the one containing $c_{2}$ describes group velocity and wavevector renormalizations associated with the distance from threshold.

\begin{figure}
\resizebox{0.5\textwidth}{!}{
\includegraphics[]{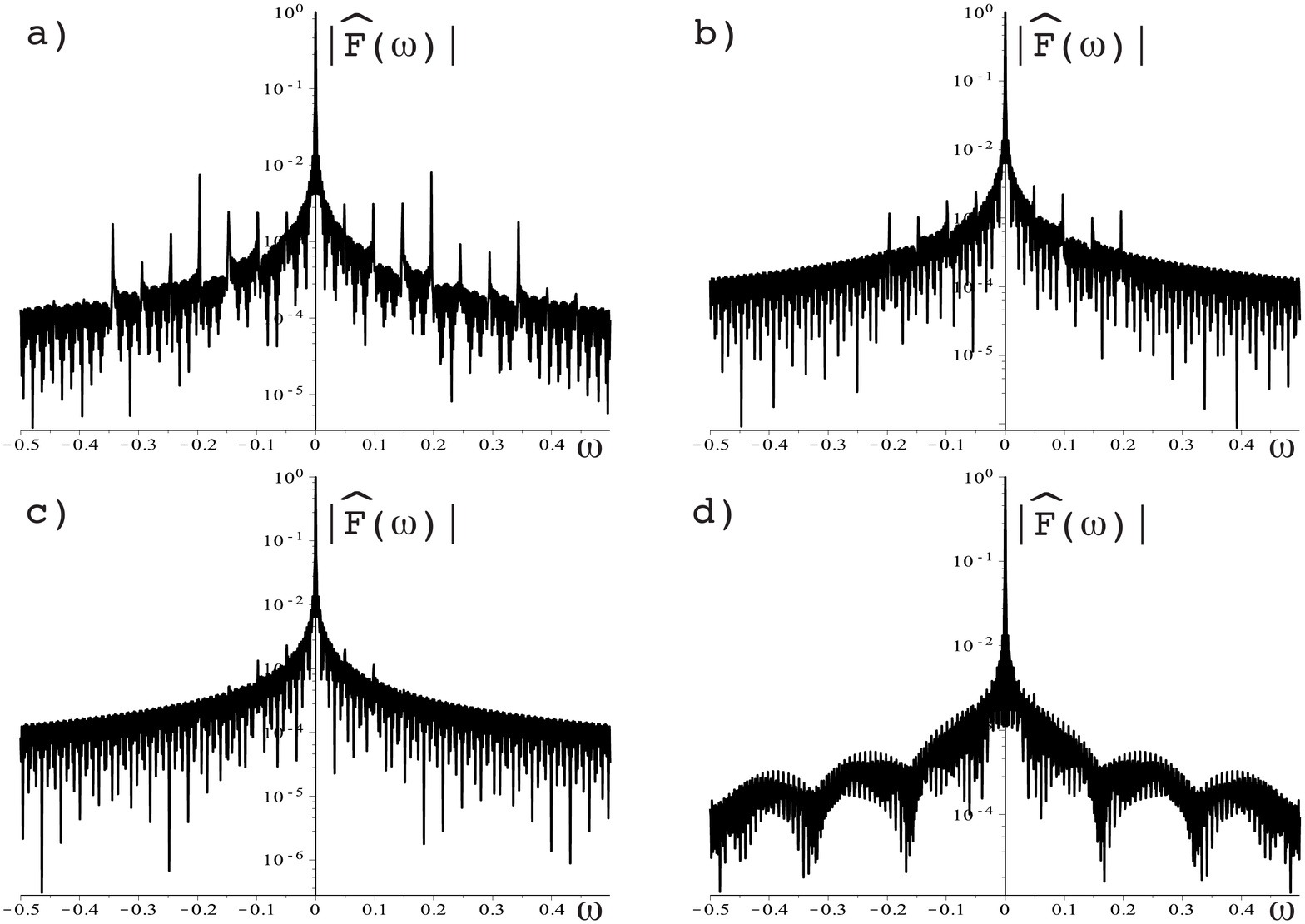}
}
\caption{Log-linear power spectrum of $F$, from eqs.~(\ref{EquationsFinales}) at different times, in the phase stability regime. $\epsilon$$=$$0.501$, $V$$=$$1$, $c_{0}$$=$$0.05+i 0.01$, $c_{1}$$=$$1.2$, $c_{2}$$=$$0$, $\mu$$=$$0.1$, $\sigma$$=$$2$, $D$$=$$1$,$\chi_{r}$$=$$3$. The time increment is $0.05$, the space increment $0.125$ and the discretization is taken on $1024$ points. The spectral components are computed in the time intervals ($\times 10^4$):  a) $[1,4]$, b) $[5,8]$, c) $[10,13]$ and d) $[57,60]$.
}
\label{fig08}
\end{figure}

We examine the stability of the spatially homogeneous solution ($S$$=$$0$, $F$$=$$\sqrt{{\tilde{\mu}}/{4 \tilde{\sigma}}}$) -- i.e., the monochromatic solution selected by the gain at threshold -- looking at its critical eigenvalue ($\lambda_{\phi}$) associated with the time translation invariance symmetry.  This eigenvalue can be shown to be expandable in power of $k$ as~\cite{inpreparation}
\begin{equation}
\lambda_{\phi}=\left(l_{2} k^2 + l_{4} k^{4} + {\cal O}(k^6) \right)+i \left(-Vk+{\cal O}(k^5)\right) \, ,
\end{equation}
with
\begin{equation}
\begin{array}{lllll}
l_{2}&=&-\epsilon^{10} &{{c_{0r}c_{1r}+c_{0i}c_{1i}}\over{c_{0r}}} &=-\epsilon^{10} c_{1r} \left(1+\alpha\beta\right)
\cr
l_{4}&=&-\epsilon^{18} &{{c_{1i}^2\left(c_{0r}^{2}+c_{0i}^{2}\right)}\over{ 2 c_{0r}^{3} \mu}} &+{\cal O}(\epsilon^{30}) \le 0
\end{array}
\end{equation}
We remark that:  (i) $l_{4}<0$, thus the small scales are damped; (ii) $l_{2}$ is the usual {\it Benjamin-Feir} phase instability control parameter \cite{BenjaminFeir1967}; $l_{2}<0$ corresponds to the stability of the single mode solution, while $l_{2}>0$ yields a phase unstable regime with possible cyclic oscillations in the optical frequency (as observed in \cite{Yacomotti2004,Furfaro2004,Tanguy2006}); (iii) The crucial difference between eqs.~(\ref{EquationsFinales}) and those of \cite{Serrat2006} lies in the existence of $c_{1i}$. Indeed, if $c_{1i}=0$ the leading term in $\Re e\{\lambda_{\phi} \}<0$ and the monochromatic solution remains stable, as in \cite{Serrat2006}, although long transients ($\lambda_{\phi}\propto{\epsilon^{10} \over L^2}$, $L$ laser length) may be mistaken for multimode dynamics (Fig.~\ref{fig08} and \cite{SerratSimulation}); (iv) $l_2>0$ if $\beta<-$$1\over\alpha$$<0$, thus the instability is controlled by $c_{1i}\over c_{1r}$ rather than $|c_1|$ (small); (v) when $l_2<0$, $\max\{\Re e\{\lambda_{\phi} \} \}$ for $k_{max}^2 \simeq -\frac{l_2}{2 l_4}$, which, substituted into $\Im m(\lambda_{\phi})$, gives a periodic oscillation of the optical frequency $\omega$ at $\Omega \simeq V k_{max}$. Since $l_2$ can approach zero, there is no lower bound for the oscillation frequency $\Omega$ and the cyclic changes in laser frequency can be as slow as desired.

\begin{figure}
\resizebox{0.5\textwidth}{!}{
\includegraphics[]{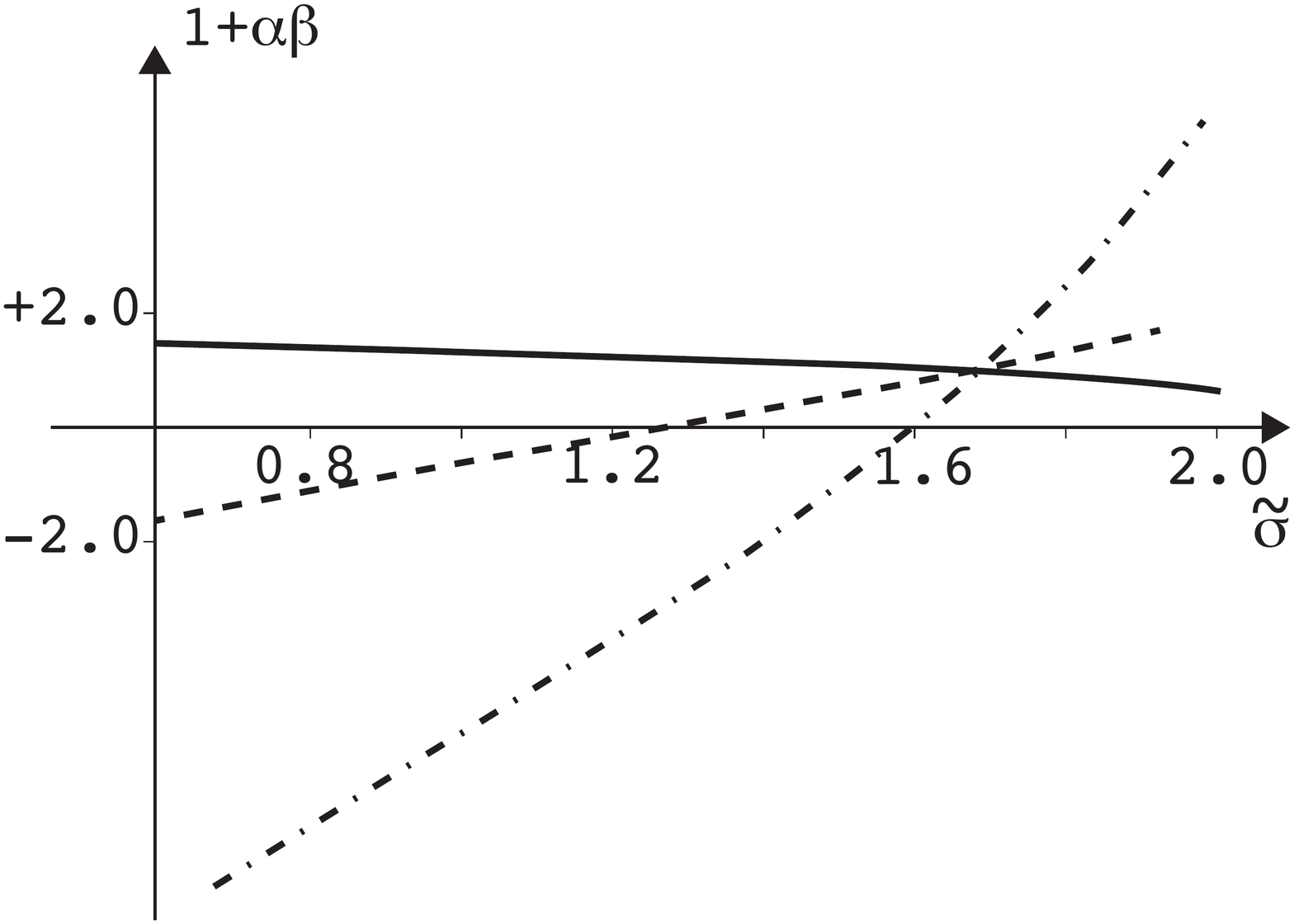}
}
\caption{Benjamin-Feir stability boundary vs. $\widetilde{\sigma}$.  The susceptibility is taken as in \cite{Balle1998}, without (solid line) and with band-gap renormalisation estimated from the experimental measurements of~\cite{Lach1991} (dashed line) or~\cite{Kulakovskii1989} (dash-dotted line).
}
\label{fig003}
\end{figure}

In order to estimate realistic physical values for $\beta$, thereby assessing the possibilities for $l_2>0$, we consider the analytical approximation for the susceptibility in MWQ lasers~\cite{Balle1998}
\begin{equation}
\chi(\omega,N)=-\chi_{0}\left[2 log\left(1-{{v}\over{u-i}}\right)-log\left(1-{{b}\over{u-i}}\right)\right] \, ,
\end{equation}
where $\chi_{0}$ is constant, $v={{N}\over{N_{pc}}}$ and $u={{\omega-{{E_{t}}\over{\hbar}}}\over{\Gamma}}$. Band-gap renormalization effects due to the screened Coulomb interaction between electrons and holes can be taken into account by renormalizing the transition energy $E_{t}$ 
\begin{equation}
E_{t}(N)=\hbar \omega_{c}-a N^{b} \Longrightarrow u=\frac{\omega-\omega_c}{\Gamma}+ p_s \left( \frac{N}{N_{pc}}\right)^{b}
\end{equation}
where $p_s$$=$$\left(\frac{aN_{pc}^b}{\hbar \Gamma} \right)$ is the bandgap shrinkage parameter~\cite{Balle1998}. The coefficients $a$ and $b$ are material-dependent and can be experimentally determined from \cite{Lach1991,Kulakovskii1989}. In Fig.~\ref{fig003} we plot $(1+\alpha\beta)$ ($\propto l_2$) vs. normalised losses $\widetilde{\sigma}$ (equivalent to changing transparency $N_{pc}$). In the absence of band-gap renormalization $(1 + \alpha\beta)>0$, thus $l_{2} < 0$ for any reasonable value of normalized losses (solid line), while the band-gap renormalization (dashed and dash-dotted lines, cf. caption) may lead to a phase unstable regime, thus proving that the phase instability is physically accessible and offering a viable interpretation for the experiments~\cite{Yacomotti2004,Furfaro2004,Tanguy2006}.

Other approximate properties of eqs.~(\ref{EquationsFinales}) may be inferred from the comparison with the CGLE, for which a vast literature exists \cite{AransonKramer}, where the phase instability appears either as pure phase turbulence or as a mixture of phase and amplitude turbulence~\cite{GLPhasePortrait}.  Although the phase gradients are strongly fluctuating in both cases, in the former the amplitude is almost constant, while in the latter its dynamics is also turbulent.  For comparison with the CGLE, even though not entirely justifiable, we perform the standard adiabatic elimination of $S$ (setting $S\simeq \widetilde{\mu}-4\widetilde{\sigma} \vert F \vert^2$ and substituting into the electric field equation), obtaining
\begin{equation}
\partial_{\widetilde{T}}F \simeq \epsilon^2 c_{0} \mu F-V \partial_{\widetilde{Z}}  F-4 \widetilde{\sigma} \epsilon^2 c_{0} \vert F \vert^2 F
+\epsilon^{10}c_{1}\partial_{\widetilde{Z}\widetilde{Z}}F+...
\end{equation}
By analogy, we then expect:  (i.) a pure phase instability regime for small $c_{0i}/c_{0r}$ and large $c_{1i}/c_{1r}$, and (ii.) an amplitude turbulent 
regime for large $c_{0i}/c_{0r}$ and small $c_{1i}/c_{1r}$.  In the pure phase instability regime, where the amplitude dynamics is enslaved to that of the phase gradients, the adiabatic elimination of the amplitude leads to the well-known Kuramoto-Sivashinsky phase equation \cite{AransonKramer} for which the number of positive Lyapunov exponents was shown to linearly increase with the system's size~\cite{Manneville}.  Thus, we expect the phase instability to act as a noise generator for the laser's electric field amplitude.

The numerical simulations of eqs.~(\ref{EquationsFinales}) are performed with a standard 4$^{\rm th}$ order Runge-Kutta algorithm in time and a 6$^{\rm th}$ order finite-difference method to approximate the spatial derivatives.  Varying the space and time increments, we have carefully checked that numerical noise does not qualitatively affect our predictions.  The simulations are performed, as usual, with periodic boundary conditions, rather than with the Fabry-Perot configuration used in the experiments~\cite{Yacomotti2004,Furfaro2004,Tanguy2006}.

\begin{figure}
\resizebox{0.5\textwidth}{!}{
\includegraphics[]{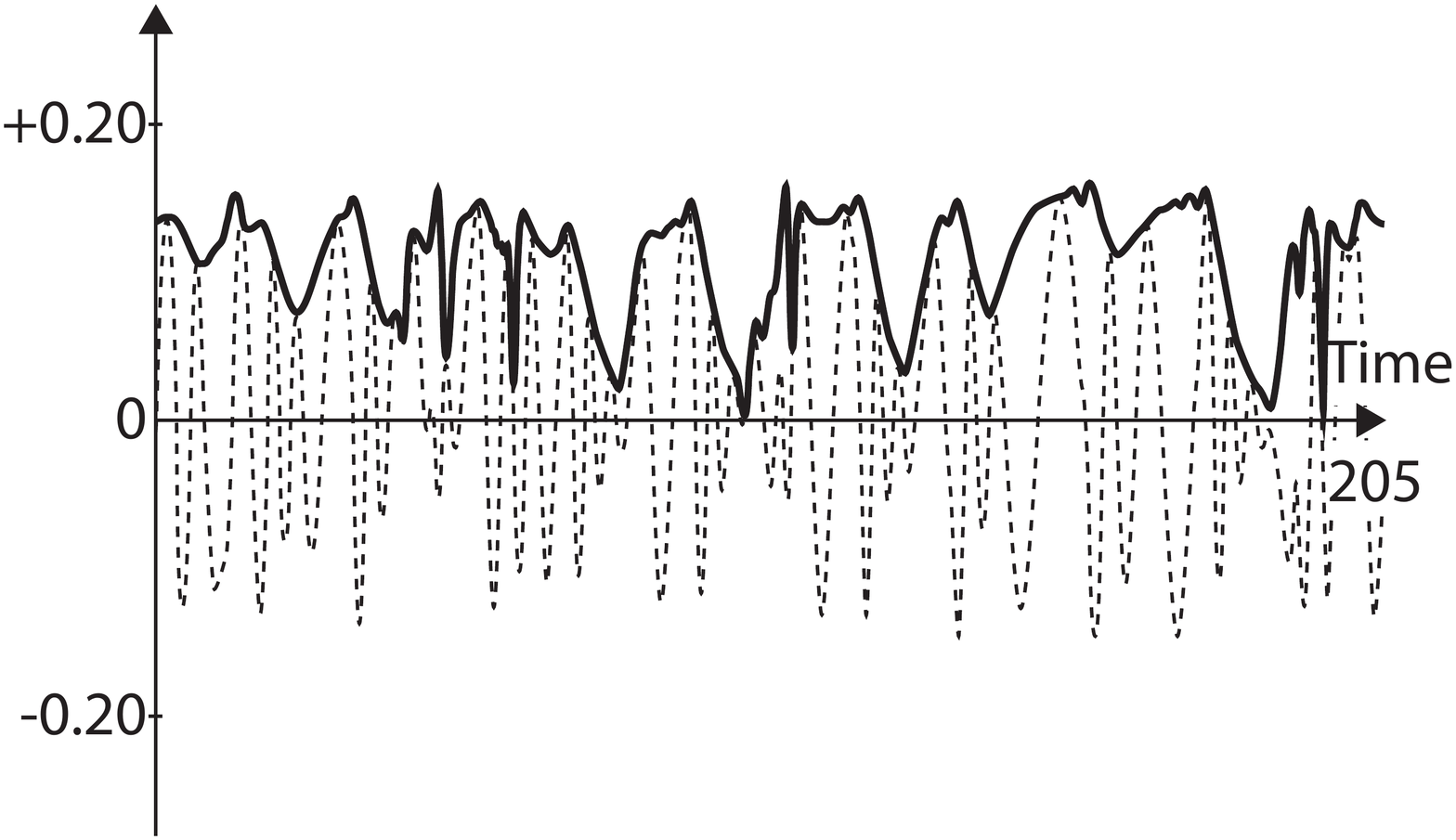}
}
\caption{Numerical simulation of eqs.~(\ref{EquationsFinales}) in the amplitude-turbulent regime. The parameters are those of Fig.~\ref{fig08}, except $c_{0}$$=$$0.02$$+$$i$$0.026$, $c_{1}$$=$$1$$-$$i$$0.9$ and $c_{2}$$=$$-i \, 10$. The continuous line represents the temporal evolution of the amplitude, at a fixed spatial position, the dashed one the real part of $F$.}
\label{fig004}
\end{figure}

\begin{figure}
\resizebox{0.5\textwidth}{!}{
\includegraphics[]{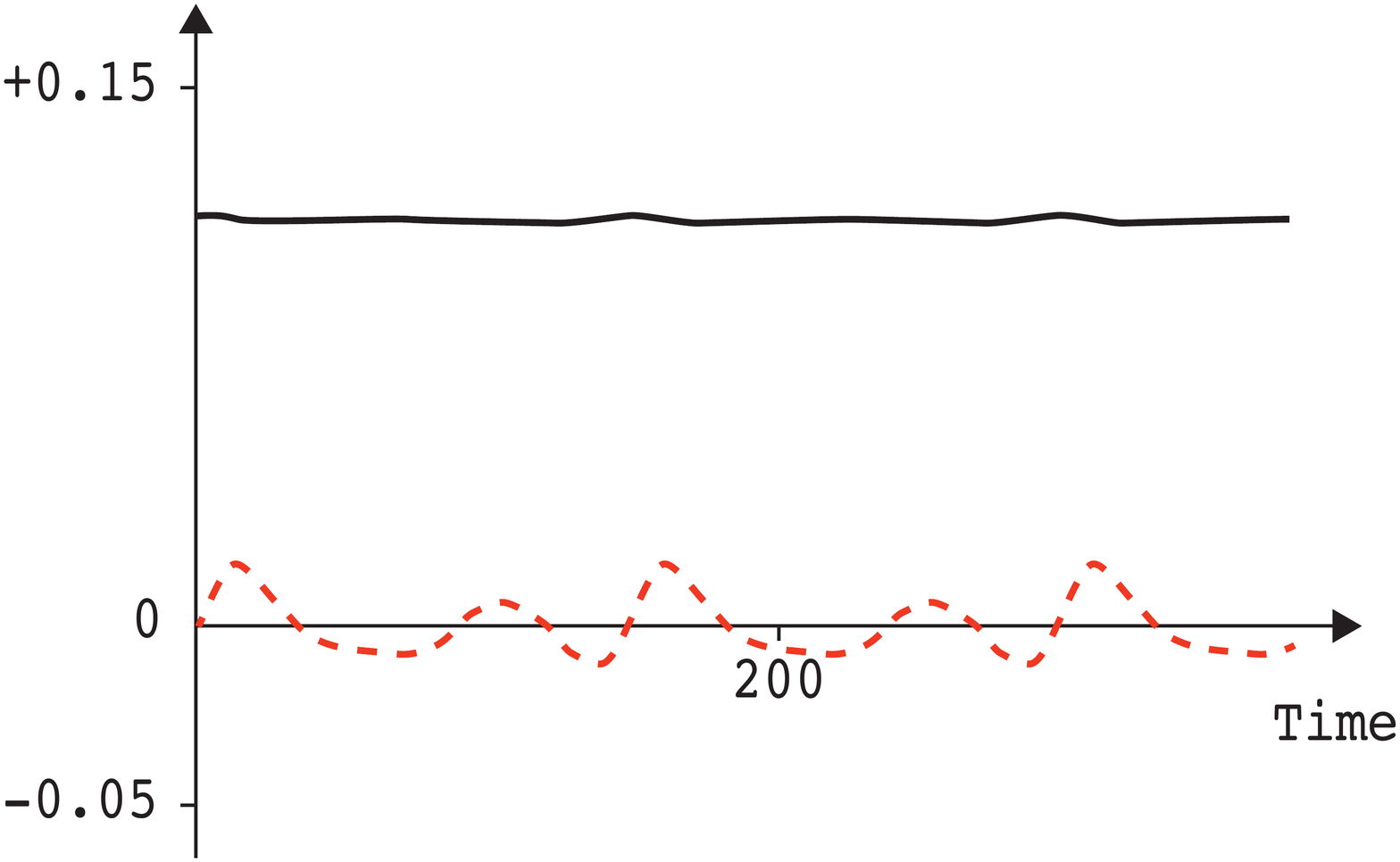}
}
\caption{Numerical simulation of eqs.~(\ref{EquationsFinales}) in the pure phase unstable regime with $c_{0}$$=$$0.05$$+$$i$$0.01$ and $c_{1}$$=$$1.2$$-$$i$$12$. The length $L$ of the numerical box is $119.3984$. The top line (black online) represents the temporal evolution of the field amplitude, at a fixed spatial position, the bottom line (red online) the time-derivative of the phase of $F$ (i.e. $\Im m\left({{F^{*}\partial_{t}F}\over{\vert F \vert^2}}\right)$). As observed experimentally, the total intensity is constant and the electric field frequency oscillation is asymmetric.}
\label{fig05}
\end{figure}

Given the long relaxation time scales expected from the phase dynamics, in order to ensure convergence in our simulations we first explore the phase-stable regime.  For the parameter values of Fig.~\ref{fig08}, the slowest phase gradient decay rate is $\lambda_{\phi}\left({{2\pi}\over{L}}\right) \sim 3 \times 10^{-6}$.  Thus, we expect, and {\bf do} observe, that the initial phase gradients vanish after a characteristic time $\tau \sim 10^6$.  On the basis of this result, in the following figures we only show predictions obtained in the asymptotic regime.  
Note that the regular aspect of the power spectrum in the asymptotic regime (Fig.~\ref{fig08}d) proves numerical noise to be negligible.  

\begin{figure}
\resizebox{0.5\textwidth}{!}{
\includegraphics[]{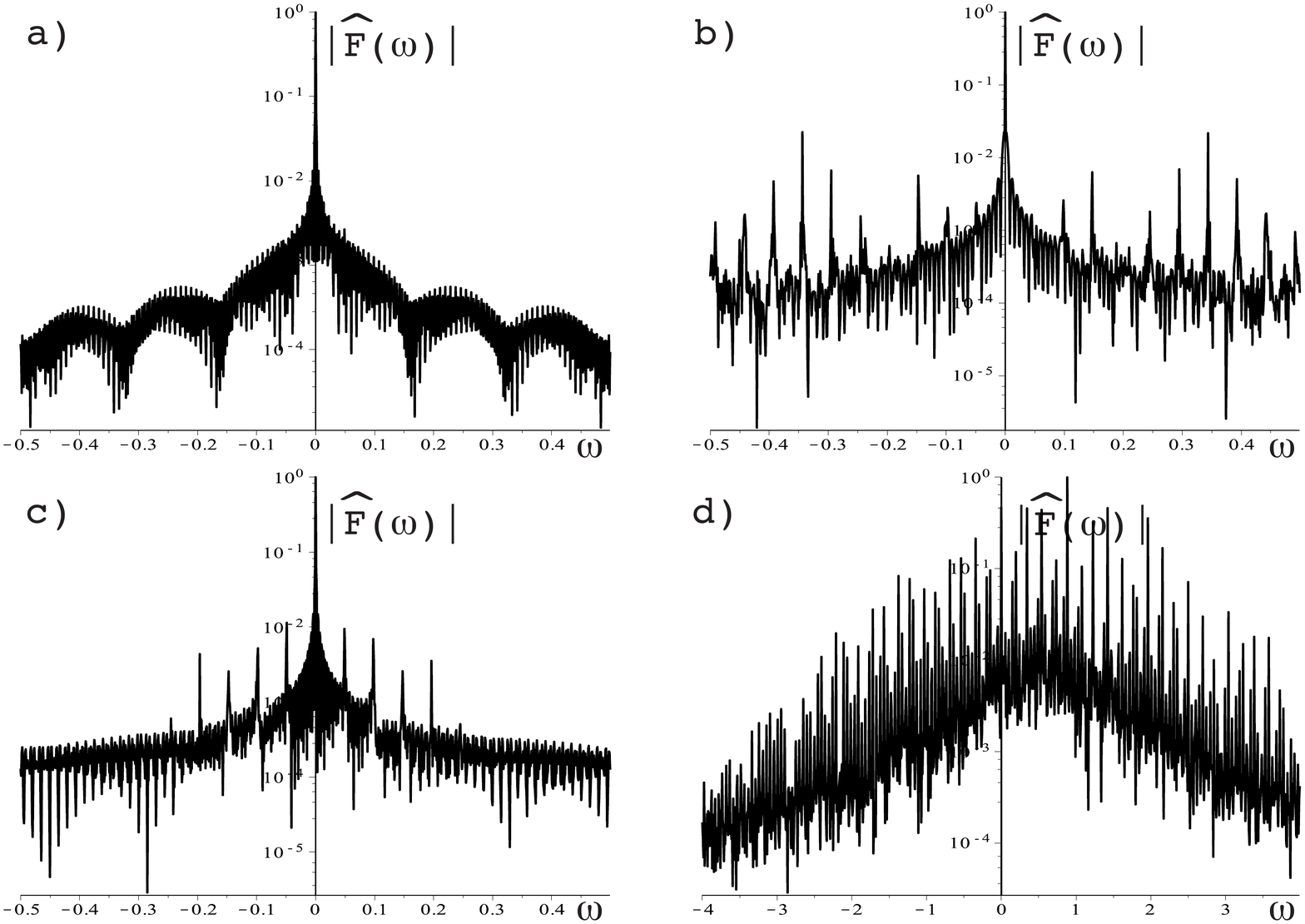}
}
\caption{Typical power spectra of $F$ in a log-linear scales obtained from the integration of eqs.~(\ref{EquationsFinales}).  a) Phase-stable regime with no added noise (aside that of the numerical scheme):  $c_{0}$$=$$0.05+i 0.01$, $c_{1}$$=$$1.2$, $c_{2}$$=$$0$. b) White noise added on space and time in the regime shown in (a).  (c) and (d) Phase-unstable regime (no added noise) -- (c) corresponds to the pure phase instability (cf. Fig.~\ref{fig05}) and (d) to amplitude turbulence (cf. Fig.~\ref{fig004}).
}
\label{fig09}
\end{figure}

By analogy with the CGLE, we associate the numerical observations of Fig.~\ref{fig004} with an amplitude turbulence regime, where not only the phase gradients but also the amplitude strongly fluctuate in space and time. The associated power spectrum is shown in Fig.~\ref{fig09}d. This parameter regime should correspond to the experimental observations obtained far from threshold, where no particular modal sequence was observed and where the total intensity oscillates irregularly \cite{privateXavier}.

Fig.~\ref{fig05} has been numerically obtained in the pure phase unstable regime (small $\alpha$, large $\beta$). The electric field frequency displays regular variations with asymmetric periodic cycling (the rise-time shorter than the fall-time). Only few modes are involved in the dynamics (Fig.~\ref{fig09}c) and the total intensity is nearly constant. These predictions are in very good qualitative agreement with the experimental observations of deterministic mode-switching \cite{Furfaro2004}, with a discrepancy in the intensity bandwidth:  in the experiment the inter-mode beatings -- if present -- could not be detected, while in our calculations they are truly absent.

Finally, we have simulated eqs.~(\ref{EquationsFinales}) in a phase stable regime but with the addition of white noise in space and time uniformly distributed between $\pm$$\zeta$$\sqrt{dt}$ where $dt$ is the time increment and $\zeta$$=$$4 \times 10^{-3}$. The aim is to compare the effect of externally injected noise to the one intrinsic to the phase instability. Although the latter involves a much narrower frequency range, they both produce multimode dynamics with somewhat differing spectral features (Fig.~\ref{fig09}b and c).

In conclusion, by computing the normal form description of a semiconductor laser bifurcation near its threshold, we have obtained a general model from which we deduce the existence of a new parameter $\beta$, proven, both analytically and numerically, to play a crucial role -- in conjunction with the well-known $\alpha$-parameter -- in the control of the phase instability.  Our numerical simulations, predicting (asymmetric) periodic oscillations in the laser frequency, are in good qualitative agreement with the experimental observations and do not need the addition of noise to obtain the deterministic dynamics seen in semiconductor laser experiments.  Even though material-related and photon noise sources are always present, and can even be strong in semiconductor devices, the ability of the phase instability to drive the system into a regular dynamical state convincingly addresses a recurrent perplexity:  how can noise, an intrinsically irregular phenomenon, induce regular dynamics? Previous answers were based either on noise playing the role of an amplitude seed transferring information, through non-linear mode coupling, from one oscillating mode to the next, or on purely deterministic, {\it ad hoc} models.  Our results show that in the phase-unstable regime the intrinsic noise may only be an additional accessory which superposes some secondary randomness onto a basically regular behavior.  Further work is needed to satisfactorily address this question.

\begin{thebibliography}{}
\bibitem{notsosmall} Hence defined, $\epsilon$ is not small ($\epsilon^1 \sim 0.5$), but in the expansions the small parameter is $\epsilon^{m} \ll 1$, since $m \gg 1$.
\end{thebibliography}
\end{document}